\documentclass[a4paper]{article}
\usepackage{graphicx}
\usepackage{amsmath}
\usepackage{url}
\usepackage{color}
\usepackage{titlesec}

\def\inststor{Institute}
\def\emailstor{Email}

\def\abststor{Abstract}
\def\kwordstor{Keywords}

\newcommand{\inst}[1]{\def\inststor{#1}}
\newcommand{\email}[1]{\def\emailstor{#1}}

\newcommand{\abst}[1]{\def\abststor{#1}}
\newcommand{\kword}[1]{\def\kwordstor{#1}}
\newcommand{\FullConference}[1]{\def\confstor{#1}}

\titleformat*{\section}{\large\bfseries}
\titlespacing*{\section}
{0pt}{2ex plus 1ex minus .2ex}{0.6ex plus .2ex}

\date{
	{\small \inststor\\
		\href{mailto:\emailstor}{\emailstor}\\
	\vspace{10pt}
	Published in the proceedings of {\it \confstor}}\\
	\abstract{\abststor}
	}

\usepackage{txfonts} 
\usepackage{upgreek}
\usepackage{bm}
\usepackage{xspace}
\usepackage[pdftex]{hyperref}

\newcommand{\refeq}[1]{Eq.~(\ref{#1})}
\newcommand{\ee}{\texorpdfstring{e$^+$e$^-$}{e+e-}\xspace}
\newcommand{\BM}{Boer--Mulders\xspace}
\newcommand{\DY}{Drell--Yan\xspace}
\newcommand{\compass}{\textsc{Compass}\xspace}
\newcommand{\hermes}{\textsc{Hermes}\xspace}
\newcommand{\djangoh}{\textsc{Djangoh}\xspace}
\renewcommand{\vec}[1]{\bm{#1}}
\newcommand{\PhT}{P_{\rm T}}
\newcommand{\PhTt}{\texorpdfstring{$\PhT$}{P\_T}\xspace}
\newcommand{\PhTv}{\vec{\PhT}}
\newcommand{\phiH}{\phi_{\rm h}}
\newcommand{\phiS}{\phi_{\rm S}}
\newcommand{\A}[2]{A_{\rm #1}^{#2}}
\newcommand{\F}[2]{F_{\rm #1}^{#2}}
\newcommand{\dif}[1]{{\rm d}{#1}\,}
\newcommand{\eps}{\varepsilon}
\newcommand{\kT}{k_{\rm T}}
\newcommand{\kTv}{\vec{\kT}}
\newcommand{\pT}{P_{\kern -0.15em \perp}}
\newcommand{\pTv}{\vec{\pT}}
\newcommand{\conv}{\mathcal{C}}
\newcommand{\hunit}{\hat{\vec{h}}}

\title{\vspace{-12mm}
        Measurements of transverse-momentum dependent effects
		in semi-inclusive DIS at \compass}
\author{Jan \textsc{Matousek}$^{1}$ on behalf of the COMPASS Collaboration}

\inst{$^{1}$Faculty of Mathematics and Physics,
		Charles University, Ke Karlovu 3, 150\,00 Praha, Czechia}

\email{jan.matousek@cern.ch}

\abst{
\compass experiment at CERN had a wide physics programme.
Its cornerstone were studies of hadron production in deep inelastic scattering (DIS), which can be interpreted in the transverse-momentum-dependent (TMD) factorisation framework, allowing to access the distributions of polarisation and transverse momentum of quarks within the nucleon in the language of TMD PDFs, and the hadronisation in terms of TMD fragmentation functions.
The data collected with a liquid hydrogen target in 2016--2017 will soon bring new information on the transverse momentum and may allow for the first time to extract the transverse polarisation of quarks within an unpolarised nucleon described by the \BM function.
The unique data collected with a transversely polarised deuteron target have already improved the knowledge of the d-quark transversity (transverse counterpart of the helicity PDF), reducing the uncertainties by a factor of 2.5 at large Bjorken~$x$, and are yet to yield a number of interesting results.
}

\FullConference{The European Physical Society Conference on High Energy Physics (EPS-HEP2025),
7--11 July 2025, Marseille, France}

\kword{DIS, SIDIS, asymmetry, Cahn, Boer--Mulders, COMPASS}

\begin{document}
\maketitle
\section{\compass}
The \compass Collaboration built a multi-purpose fixed-target setup in the North Area of CERN, 
which was taking data between 2002 and 2022~\cite{compass:prop1,compass:prop2}.
It occupied the place of the former SMC, NMC and EMC experiments on the unique M2 beamline
capable of delivering both positive and negative secondary hadron beams or tertiary muon beams
longitudinally polarised by weak decay.
The Collaboration has run a wide programme, comprising semi-inclusive measurements of deep inelastic scattering (DIS) of muons on longitudinally or transversely polarised targets, the first measurement of Drell--Yan process with transversely polarised target, measurements of hard exclusive processes on liquid hydrogen target, hadron spectroscopy and studies of chiral dynamics.
After 20 years of data collection, the experiment is now in the analysis phase. For example, recently, the polarised Drell--Yan analysis has been concluded~\cite{COMPASS:2023vqt} and the results for the differential cross-section are being finalised~\cite{Andrieux:2024:04}. Publications on hard exclusive $\uppi^0$ muoproduction~\cite{COMPASS:2024hvm} and strange meson spectroscopy~\cite{COMPASS:2025wkw} have been submitted for publication.
In this text, the recent results and progress in the analysis of DIS measurements are summarised.

\section{Hadron production in DIS}
The cross section for a semi-inclusive measurement of hadron production in DIS off a transversely polarised target $\upmu {\rm N}^\uparrow \rightarrow \upmu^\prime {\rm h X}$ is in terms of the usual variables $x$, $y$ and $Q^2$, 
the fraction $z$ of the energy transferred to the hadron, its transverse momentum \PhTt, and the azimuthal angles $\phiH$ of the hadron momentum and $\phiS$ of the target polarisation in the gamma--nucleon system~\cite{Bacchetta:2006tn}:   
\begin{align} \label{eq:xsec}
    &\frac{\dif{\sigma}}{\dif{x} \dif{y} \dif{z} \dif{\PhT^2} \dif{\phiS} \dif{\phiH}}
        = \frac{\alpha}{xyQ^2}\frac{y^2}{2(1-\eps)} \biggl( 1 + \frac{\gamma^2}{2x} \biggr)
            \bigl( \F{UU,T}{} + \eps \F{UU,L}{} \bigr)
            \nonumber \\ &~
            \times \biggl\{ 1
            + \sqrt{2\eps(1+\eps)} \A{UU}{\cos\phiH}\cos\phiH
            + \eps \A{UU}{\cos2\phiH}\cos2\phiH
            + \lambda \sqrt{2\eps(1-\eps)} \A{LU}{\sin\phiH}\sin\phiH
            \nonumber \\ &\qquad
            + S_{\rm T} \bigl[ \A{UT}{\sin(\phiH-\phiS)}\sin(\phiH-\phiS)
                            + \eps \A{UT}{\sin(\phiH+\phiS)}\sin(\phiH+\phiS)
                            + \eps \A{UT}{\sin(3\phiH-\phiS)}\sin(3\phiH-\phiS)
                            \nonumber \\ &\qquad\qquad\quad
                            + \sqrt{2\eps(1+\eps)} \A{UT}{\sin\phiS}\sin\phiS
                            + \sqrt{2\eps(1+\eps)} \A{UT}{\sin(2\phiH-\phiS)}\sin(2\phiH-\phiS)
                            \bigr]
            \nonumber \\ &\qquad
            + \lambda S_{\rm T} \bigl[  \sqrt{1-\eps^2} \A{LT}{\cos(\phiH-\phiS)}\cos(\phiH-\phiS)
                            + \sqrt{2\eps(1-\eps)} \A{LT}{\cos\phiS}\cos\phiS
                            \nonumber \\ &\qquad\qquad\quad
                            + \sqrt{2\eps(1-\eps)} \A{LT}{\cos(2\phiH-\phiS)}\cos(2\phiH-\phiS)
                            \bigr]
        \biggr\},
\end{align}
where $\gamma = 2Mx/Q$, $M$ being the mass of the target nucleon,
$\eps = \bigl(1-y-\frac{1}{4}\gamma^2\eps^2\bigr)\big/\bigl(1-y+\frac{1}{2}y^2+\frac{1}{4}\gamma^2y^2\bigr)$,
$\lambda$ is the longitudinal beam polarisation, $S_{\rm T}$ is the transverse target nucleon polarisation,
$F$ are structure functions and $A$ are asymmetries, both being in general functions of $x$, $z$, $\PhT$ and $Q^2$.
The asymmetries -- the amplitudes of the azimuthal modulations in \refeq{eq:xsec} -- are ratios of structure functions, e.g., $\A{UU}{\cos\phiH} = F_{\rm UU}^{\cos\phiH}/(F_{\rm UU,T} + \eps F_{\rm UU,L})$.
If the transverse momentum is small, $\PhT \ll z Q$, each structure function can be interpreted in the transverse-momentum-dependent (TMD) factorisation framework as a convolution of a TMD PDF $f(x,\kT^2,Q^2)$ and fragmentation function (FF) $D(z,\pT^2,Q^2)$, which depend on the intrinsic quark transverse momentum within the nucleon $\kT$ and the transverse momentum of the hadron $\pT$ obtained in hadronisation, weighted with a function $w(\kTv\cdot\PhTv, \pTv\cdot\PhTv, \kTv\cdot\pTv)$,
\begin{equation}
    \conv[w fD] = x \sum_q e^2_q \int \dif{^2\kTv} \dif{^2\pTv} \delta^{(2)}(z \kTv + \pTv - \PhTv)
		w \, f^q \, D^{q \to h}.
\end{equation}
\vspace{-15pt}

Measurements of the amplitudes of the azimuthal modulations in \refeq{eq:xsec} give access to a variety of TMD PDFs, describing correlations between the nucleon and quark polarisation and transverse momentum~\cite{Bacchetta:2006tn}.
Up to the order $1/Q$ (sub-leading twist) and denoting `...' terms vanishing in a Wandzura--Wilczek-type approximation (neglecting terms related to qgq correlators~\cite{Bastami:2018xqd}), we have:
\begin{gather} \label{eq:TMDs}
    \F{UU,T}{} = \mathcal{C}\left[ f_1 D_1 \right],
    \quad
    \F{UU,L}{} = 0,
	\quad
	\F{UU}{\cos2\phiH} = \mathcal{C}\left[
		\frac{2(\hunit\cdot\kTv)(\hunit\cdot\pTv) - \kTv\cdot\pTv}{MM_{\rm h}}
		h_1^\perp H_1^\perp \right],
	\nonumber\\
	\F{LU}{\sin\phiH} = \frac{2M}{Q}\mathcal{C}\left[...\right],
	\qquad
	\F{UU}{\cos\phiH} = \frac{2M}{Q}\mathcal{C}\left[
		- \frac{\hunit\cdot\kTv}{M} f_1 D_1
		- \frac{(\hunit\cdot\pTv) \kT^2}{M^2M_{\rm h}}
		h_1^\perp H_1^\perp + ...\right],
	\nonumber\\
	\F{UT}{\sin(\phiH-\phiS)} = \conv \left[
		-\frac{\hunit \cdot \pTv}{M} f_{\rm 1T}^\perp D_1 \right],
    \qquad
	\F{UT}{\sin(\phiH+\phi_S)} = \conv \left[
		-\frac{\hunit \cdot \kTv}{M_h} h_1 H_1^\perp \right],
\end{gather}
where we used a unit vector $\hunit = \PhTv/\PhT$ and denoted $M_{\rm h}$ the mass of the produced hadron.
The structure function $\F{UU,T}{}$ gives access to the TMD PDF of unpolarised quarks within an unpolarised nucleon $f_1$ and the TMD FF $D_1$, describing the hadronisation of an unpolarised quark.

In an unpolarised nucleon, quarks may be transversely polarised with respect to their transverse momentum, as described by the \BM TMD PDF $h_1^\perp$~\cite{Boer:1997nt}. This would lead to $\A{UU}{\cos2\phiH}$ asymmetry due to the azimuthal-angle-dependent hadronisation of transversely polarised quarks described by the Collins TMD FF $H_1^\perp$~\cite{Collins:1992kk}.
The $h_1^\perp$ also contributes to the asymmetry $\A{UU}{\cos\phiH}$, 
which is expected to receive the dominant contribution from a kinematic effect arising from a non-zero $\langle k_T \rangle$ predicted by Cahn~\cite{Cahn:1978se}. 
$\A{LU}{\sin\phiH}$ arises from purely sub-leading twist effects (qgq correlators).

In a transversely polarised nucleon, the asymmetry $\A{UT}{\sin(\phiH-\phiS)}$ is induced by the Sivers TMD PDF $f_{\rm 1T}^\perp$, which describes the correlation between the quark transverse momentum and the nucleon transverse spin~\cite{Sivers:1989cc}, 
while the asymmetry $\A{UT}{\sin(\phiH+\phiS)}$ arises from the transversity $h_1$ (the transverse counterpart of the helicity PDF $g_1$) coupled to the Collins function $H_1^\perp$.

Measurements of these structure functions (or asymmetries) in DIS, combined with results from the \DY process and semi-inclusive hadron production in \ee annihilation, allow phenomenologists to extract the TMD PDFs and FFs, which become benchmarks for lattice QCD calculations and models, and play a role in the interpretation of other measurements.

Although for $\upmu {\rm N} \rightarrow \upmu^\prime {\rm h X}$ at $Q^2>1~({\rm GeV}/c)^2$ DIS is the leading process, there is an important background: the decay of diffractively produced vector mesons V, 
$\upmu {\rm N} \to \upmu^\prime {\rm N^\prime V}$, ${\rm V \to h^+h^-}$.
The production of $\uprho$, decaying to $\uppi^+\uppi^-$, has the largest cross section, while $\upphi \to {\rm K^+K^-}$ is important if the production of kaons is studied.
The contamination of the DIS hadron sample is small in most of the phase space, but can reach 50\% at large $z$ and low $\PhT$ and $Q^2$. 
In addition, the hadrons exhibit large azimuthal modulations $\A{UU}{\cos\phiH}$ and $\A{UU}{\cos2\phiH}$~\cite{Agarwala:2019kqu}, which need to be corrected (unlike $\A{UT}{f}$).

The TMD framework does not account for the radiative effects of QED.
In particular, a real photon emission not only distorts the reconstructed $x$ and $Q^2$, which can be corrected rather easily, but also impacts the hadron variables $z$, \PhTt and $\phiH$, calling for radiative corrections\footnote{For $\A{UT}{f}$, the radiation just dilutes the asymmetry. In \compass approach, it is included in the target dilution factor.}.

\section{Experimental results}
The \PhTt-dependent multiplicity of hadrons produced in DIS off an isoscalar target at \compass~\cite{COMPASS:2017mvk} became the cornerstone of global extractions of $f_1^q(x,\kT^2,Q^2)$ and $D^{q \to h}(z, \pT^2, Q^2)$, such as in Ref.~\cite{Bacchetta:2024qre}.
\hermes~\cite{HERMES:2012uyd} contributed with flavour sensitivity thanks to the use of hadron identification and H and D targets.
This year, \compass published the collinear (\PhTt-integrated) multiplicity of identified $\pi^\pm$ and K$^\pm$ on the liquid hydrogen target~\cite{COMPASS:2024gje}, with a more robust treatment of the QED radiative effects using the \djangoh Monte Carlo generator~\cite{Aschenauer:2013iia, Djangoh:github}, allowing to correct the distributions of the hadron variables $z$ and $\PhT$. The work on $\PhT$-dependence is ongoing~\cite{Benesova:iwhss2025}.

\begin{figure}[tb]
\centering\vspace{-10pt}
\includegraphics[width=0.7\textwidth]{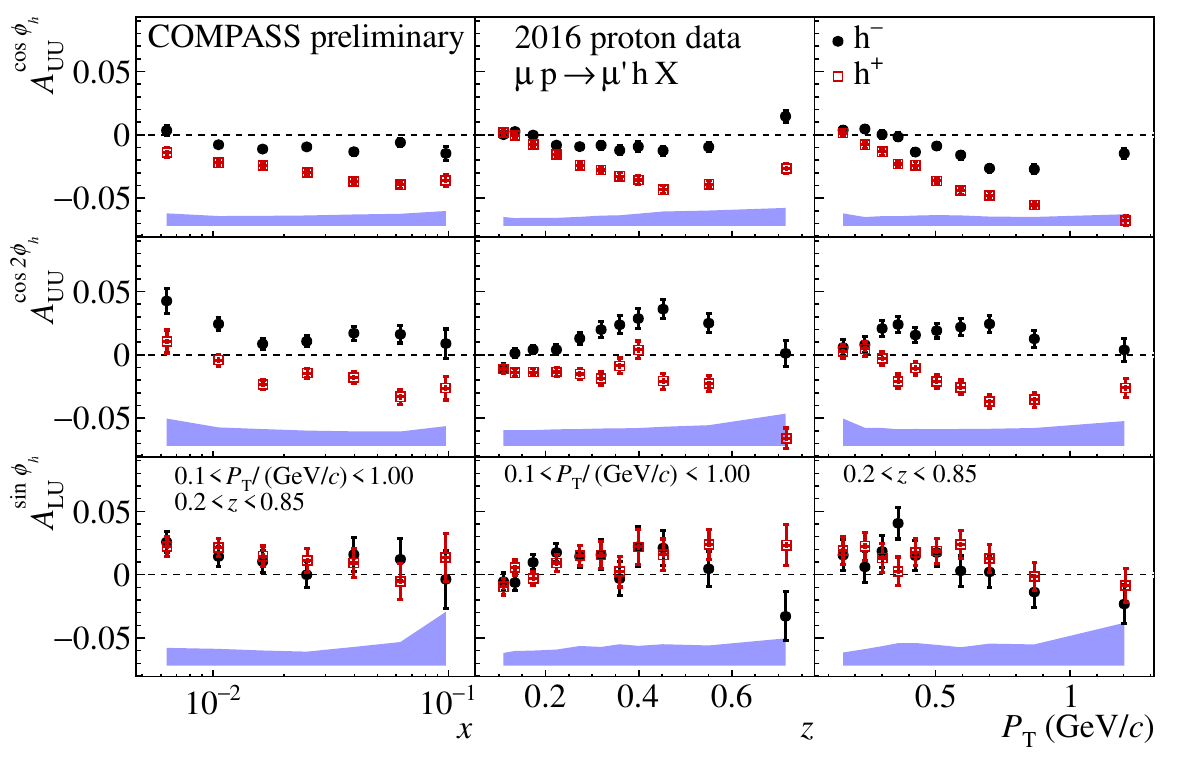}
\caption{\label{fig:ua:results}
	The final amplitudes of azimuthal modulations of positive and negative hadrons (h$^\pm$). 
    The error bars denote statistical uncertainties,
    the bands (common for h$^+$ and h$^-$) show the systematic ones.}
\end{figure}

\begin{figure}[tb]
\centering\vspace{-10pt}
\includegraphics[width=0.7\textwidth]{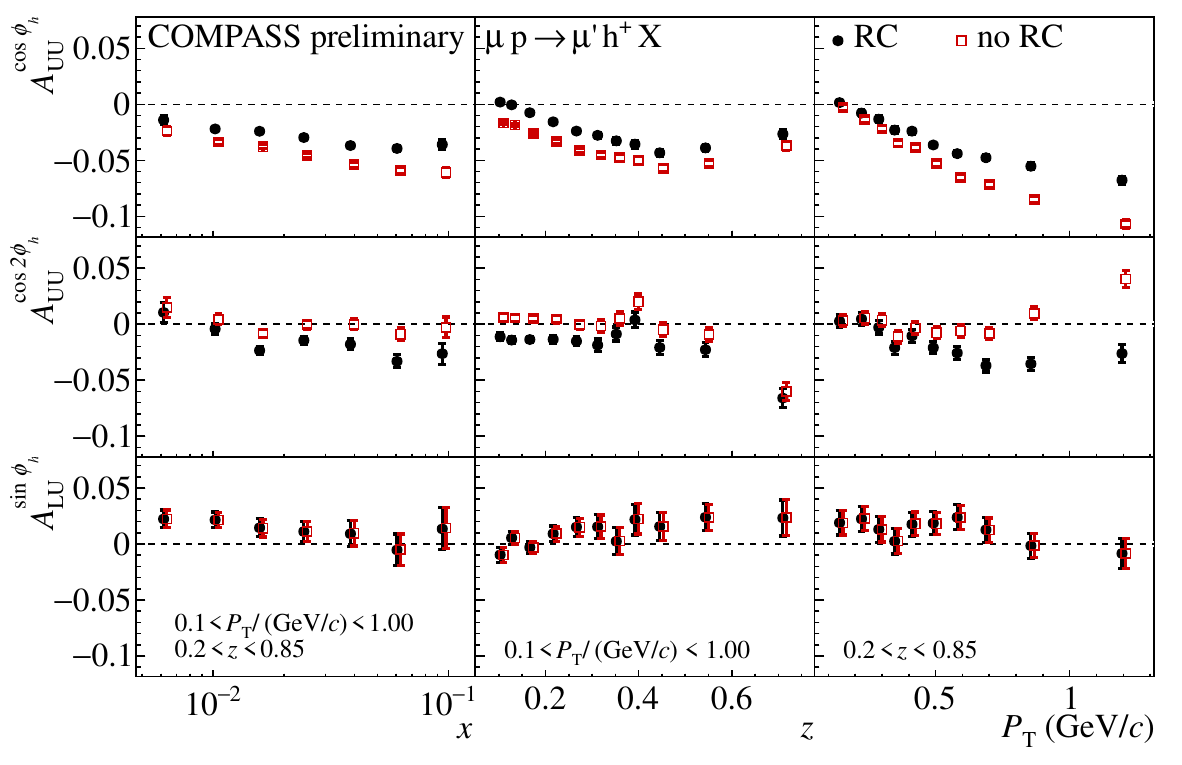}
\caption{\label{fig:ua:rceffect}
	A comparison of the amplitudes for h$^+$ before (no RC) and after (RC)
    radiative corrections based on \djangoh. The radiative corrections for h$^-$ are the same
    as for h$^+$~\cite{Benesova:2024cfk}.}
\end{figure}

The target-polarisation-independent asymmetries $\A{UU}{\cos\phiH}, \A{UU}{\cos2\phiH}$ and $\A{LU}{\sin\phiH}$ were shown to be non-zero by several experiments, e.g. by \hermes on hydrogen and deuterium~\cite{HERMES:2012kpt}, 
\compass on an isoscalar target~\cite{Adolph:2014pwc} or {\sc Clas} on hydrogen~\cite{CLAS:2021opg}.
However, their interpretation was difficult because of the interplay of different contributions. In particular, the attempt to extract the \BM function and $\langle \kT \rangle$ from the asymmetries and \PhTt-distributions faced challenges in finding a simultaneous description of all the observables~\cite{Barone:2015ksa}.
At the time, they were attributed to higher-twist effects.
In Fig.~\ref{fig:ua:results}, we show recent preliminary results on the asymmetries measured on the liquid hydrogen target~\cite{Benesova:2024cfk}, which were corrected for the background from diffractive vector mesons and for QED radiative effects.
The latter are sizeable, as shown in Fig.~\ref{fig:ua:rceffect}.
The measurement differential in $x, z, \PhT$ and $Q^2$ -- crucial to disentangling the different contributions -- is being finalised.

\begin{figure}[tb]
\centering\vspace{-5pt}
\includegraphics[width=0.9\textwidth]{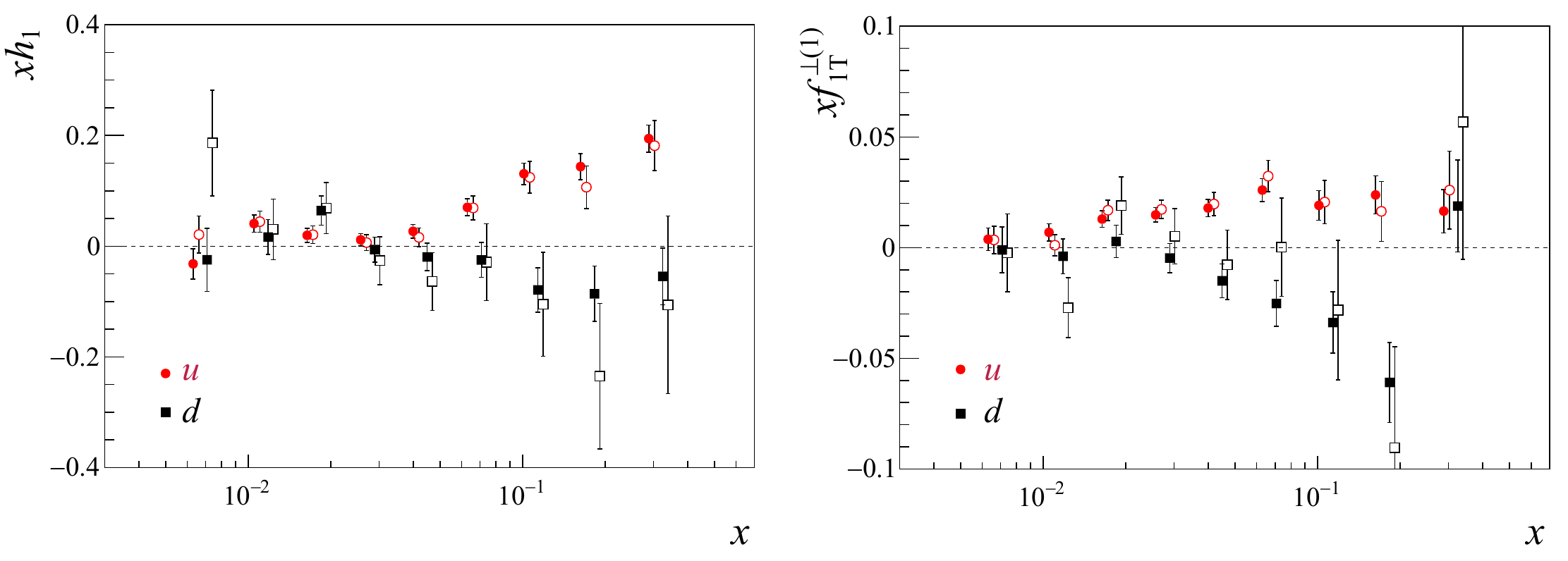}
\caption{\label{fig:ext}
	Point-by-point extractions of transversity (left) and of the first moment
    of the Sivers function (right) using \compass data available before 2022
    (open points) and including 75\% of the new deuteron data (closed points)
    demonstrate the impact the new results would have on global fits~\cite{COMPASS:2023vhr}.}
\end{figure}

The Sivers and Collins transverse spin asymmetries $\A{UT}{\sin(\phiH-\phiS)}$ and $\A{UT}{\sin(\phiH+\phiS)}$ were first measured to be non-zero on protons by \hermes~\cite{hermes:2004} and compatible with zero on deuterons by \compass~\cite{compass:2005t}.
Later, more data were collected, e.g., on protons by \compass~\cite{Adolph:2014zba}. 
However, the uncertainties for deuterons remained relatively large, leading to a large uncertainty especially of the d-quark transversity in global fits (e.g. Ref.~\cite{Gamberg:2022kdb}).
In 2022, \compass collected a large sample of DIS events with transversely polarised deuteron ($^6$LiD) target, shrinking the uncertainties on the asymmetries up to three times~\cite{COMPASS:2023vhr}. The extractions shown in Fig.~\ref{fig:ext} demonstrate the expected impact on the knowledge of d-quark distributions.
Work is ongoing on releasing the full dataset and more observables, such as asymmetries of hadron pairs~\cite{Asatryan:2024cyt} or identified hadrons.
\begin{figure}[tb]
\centering\vspace{0pt}
\includegraphics[width=0.7\textwidth]{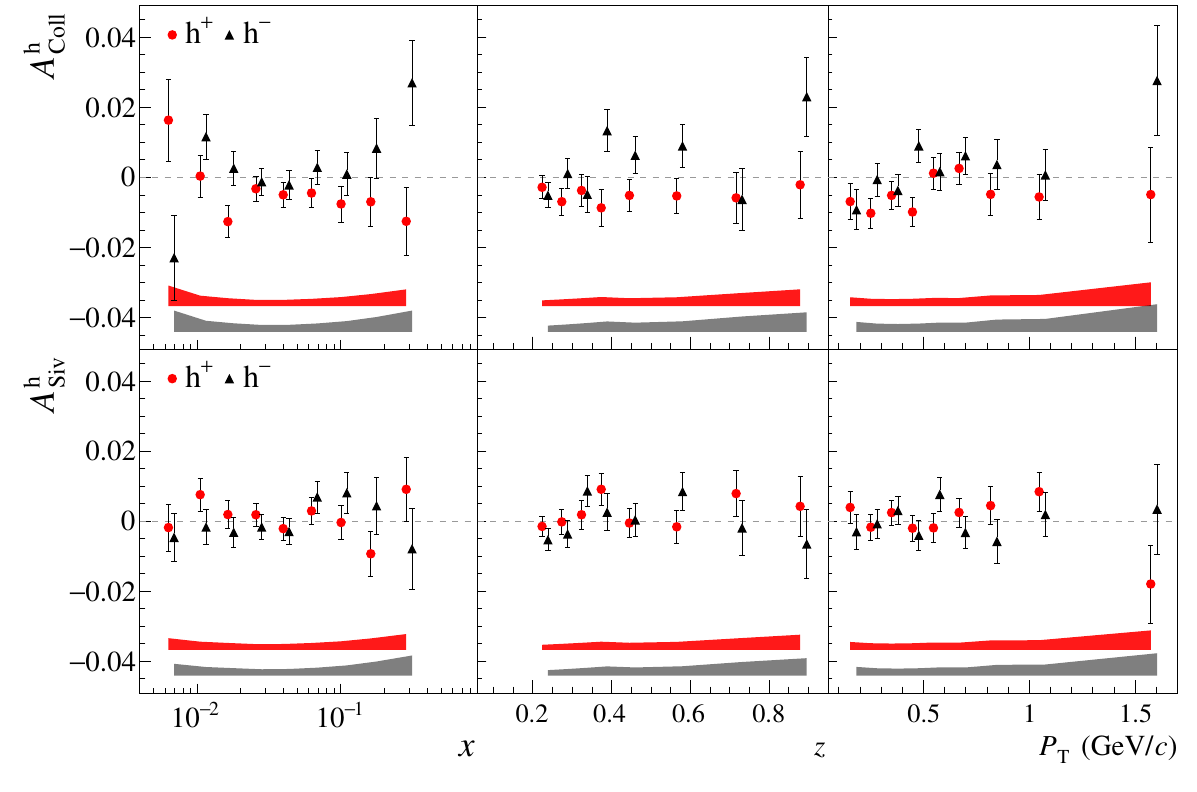}
\caption{\label{fig:ColSiv}
The Collins and Sivers asymmetries $\A{UT}{\sin(\phiH+\phiS)}$ and  $\A{UT}{\sin(\phiH-\phiS)}$ measured on transversely polarised deuterons.
    The error bars denote statistical uncertainties,
    the bands show the systematic ones.~\cite{COMPASS:2023vhr}.}
\end{figure}

\section{Conclusions and outlook}
Deep inelastic scattering (DIS) is a powerful tool for investigating the transverse-momentum- and transverse-spin-dependent nucleon structure.
Intense work is ongoing on the datasets collected in 2016--2017 with the liquid hydrogen target and in 2022 with the transversely polarised deuteron ($^6$LiD) target by the \compass Collaboration at CERN.
Both are about to yield interesting results. The hydrogen one because of advances in analysis methods such as the treatment of the background from diffractive vector mesons and radiative corrections, and the other one because of the absolute uniqueness of polarised deuteron data.
\section*{Acknowledgements}
The speaker has been supported by the FORTE project, {\small \tt CZ.02.01.01/00/22 008/0004632}, from Czech MEYS, co-funded by the EU, and by Charles University grant {\small \tt PRIMUS/22/SCI/017}.
\providecommand{\href}[2]{#2}\begingroup\raggedright\endgroup

\end{document}